# Extracting possibly representative COVID-19 Biomarkers from X-Ray images with Deep Learning approach and image data related to Pulmonary Diseases

Ioannis D. Apostolopoulos[1], Aznaouridis I. Sokratis[2], Tzani A. Mpesiana[3],

[1] Department of Medical Physics, School of Medicine, University of Patras, 26504 Patras, Greece
[2] Department of Computer Engineering and Informatics, University of Patras, 26504 Patras, Greece
[3] Computer Technology Institute and Press "Diophantus", 26504 Patras, Greece

Correspondence: Ioannis D. Apostolopoulos, ece7216@upnet.gr

**Abstract.**
In this study, the problem of automatically classifying pulmonary diseases, including the recently emerged COVID-19, from X-Ray images, is considered. While the spread of COVID-19 is increased, new, automatic, and reliable methods for accurate detection are essential to reduce the exposure of the medical experts to the outbreak. X-Ray imaging, although limited to specific visualizations, may be helpful for the diagnosis. Deep Learning has proven to be a remarkable method to extract massive high-dimensional features from medical images. Specifically, in this paper, the state-of-the-art Convolutional Neural Network called Mobile Net is employed and trained from scratch to investigate the importance of the extracted features for the classification task. A large-scale dataset of 3905 X-Ray images, corresponding to 6 diseases is utilized for training MobileNet v2, which has been proven to achieve remarkable results in related tasks. The results suggest that training CNNs from scratch may reveal vital biomarkers related but not limited to the COVID-19 disease, while an overall classification accuracy of the seven classes reaches 87.66%. Besides, this method achieves 99.18% accuracy, 97.36% Sensitivity, and 99.42% Specificity in the detection of COVID-19.

## 1   Introduction

The Coronavirus (COVID-19) is perhaps the greatest challenge of mankind in the 21st century. The development of the disease, its transmission, and the increased mortality in a number of countries, make it imperative to develop treatment, but also to protect health care and society from the transmission of the disease.

Therefore, remote control of the disease, including diagnosis, early quarantine, and follow-up, is imperative. Artificial intelligence can contribute to the above perspectives. Recent studies claim to achieve remarkable results regarding the automatic detection of the disease from thoracic X-Ray scans [1-3]. Although the research is limited due to the absence of large scale image data, the first results are encouraging and necessitate further investigation and research.

Although the diagnosis is increasingly becoming a rapid process, the financial issues arising from the cost of diagnostic tests concern both states and patients, especially in countries with private health systems, or restricted access health systems due to prohibitive prices.

During the last months, there has been an increase in publicly available patient data, including X-Ray images. Possible patterns and knowledge mined from the X-Ray scans may constitute a possible pipeline for the diagnosis of COVID-19.

The development of deep learning applications over the last five years enable the researchers to perform a rapid and deep analysis on the X-Ray scans. Deep Learning is a combination of Machine Learning methods mainly focused on the automatic feature extraction and classification from images, while its applications are broadly met in medical image detection, segmentation, and classification tasks. Machine learning and Deep Learning have become established disciplines in applying artificial intelligence to mine, analyze, and recognize patterns from data. Reclaiming the advances of those fields to the benefit of clinical decision making and computer-aided systems is increasingly becoming nontrivial, as new data emerge [4].

Deep Learning is a learning method wherein deep Convolutional Neural Networks (CNN) are utilized for automatic mass feature extraction, achieved by the process called convolution [5]. Each layer involves a transformation of the data into a higher and more abstract level. Higher layers (i.e., deep layers) of portrayal enhance parts of the information that are significant for segregation and smother unimportant attributes. Due to the unlimited parameters mined during this process, several methods have been proposed to achieve dimensionality reduction, such as Pooling [5].

Motivated by the recent and relative research, in this study, we focus on circumventing two vital issues arisen in the detection of COVID-19 from X-Ray scans. The first issue is related to the methodology of the experimental setups. In essence, the researches have demonstrated that the detection of COVID-19 is achievable, but this conclusion derives from an analysis based on incomplete data. The models proposed are powerful in classifying images between only three classes (viral and bacterial pneumonia, COVID-19, normal). This, unfortunately, does not in demonstrated the existence of a clear fingerprint of the Coronavirus in the X-Ray images, firstly due to the insignificant database size, and secondly due to the fact that the fingerprints of other pulmonary diseases have not been compared. The second issue is related to the flaws of Deep Learning and is often referred to as the issue of interpretability [6]. In short, the algorithm is not transparent, thereby a radiologist cannot supervise and know which factors or indices the model was based on to predict.

To further evaluate the methodology of Deep Learning, we perform an experiment utilizing six of the most common pulmonary diseases, including the COVID-19. In this way, the capabilities of the method in distinguishing between the various diseases is evaluated. Besides, the dataset of the particular experiment is significant, including approximately 450 cases of COVID-19. To contribute to the latter referred issue (i.e., the interpretability), we perform three different experiments altering the mining methods to inspect the variance of the extracted features. Specifically, the state-of-the-art CNN called Mobile Net (v2) is employed to extract features from the images in three different ways, as follows: a) training from scratch, b) feature extraction via transfer learning (or of-the-self-features), and c) hybrid feature extraction via fine-tuning. Those methods are explained in Section 2.2.

Due to the absence of a complete X-Ray dataset containing not only common pneumonia or other diseases, but also cases of diagnosed COVID-19, the final dataset of this experiment is a combination of X-Rays corresponding to common pulmonary diseases recorded during the last years and confirmed COVID-19 cases recorded the last months.

The results of the present research further enhance the research to date. In particular, it is highlighted that with the strategy of training from scratch, the CNN succeeds in mining significant image features, discovered solely in the particular X-Ray images. Based on these characteristics (features), 88% accuracy in classification of the relative diseases and ~ 99% accuracy in diagnosis of COVID is achieved. This may prove that these features are Biomarkers and need further analysis, as they may be gene or other signatures.

## 2  Methods

### 2.1  Dataset of the study

#### 2.1.1  COVID-19 X-Ray images

For the creation of the dataset, the research focused on obtaining X-Rays corresponding to confirmed cases infected by the virus SARS-COV-2. Through extensive research, a collection of 455 well-visualized, confirmed pathological X-Ray images was created. The final collection includes selected X-Rays from a Github repository created by Dr. Cohen [7], and publically available medical image repositories, such as the Radiological Society of North America (RSNA), Radiopaedia, and the Italian Society of Medical and Interventional Radiology (SIRM). The latter association released a publically available COVID-19 dataset [8], which was also incorporated.

#### 2.1.2 Common Bacterial and Viral Pneumonia X-Ray images

To train and evaluate the classification method in more complex conditions, a collection of conventional bacterial and viral pneumonia X-Ray scans was added to the dataset. This collection is available on the Internet by Kermany et al. [9]. A selection of 910 related X-Ray images was incorporated into the dataset.

#### 2.1.3 Pulmonary diseases detected from X-Ray scans

It is impossible to investigate the performance of any classification method in detecting the COVID-19 disease unless other pulmonary diseases are incorporated. For this reason, the final dataset includes selected X-Ray scans corresponding to other pulmonary abnormalities.

The National Institutes of Health (NIH) X-Ray dataset was exploited, which is referred to as NIH dataset for the particular experiment, and comprises 112.120 frontal-view X-ray images of 30.805 unique patients with the text-mined fourteen disease image labels [10].

Those images are extracted from the clinical PACS database at the National Institutes of Health Clinical Center in America. The corresponding diseases were mined from the associated radiological reports using natural language processing. The labels contain fourteen common thoracic pathologies include Atelectasis, Consolidation, Infiltration, Pneumothorax, Edema, Emphysema, Fibrosis, Effusion, Pneumonia,

Pleural thickening, Cardiomegaly, Nodule, Mass and Hernia. This dataset is significantly more representative of the real patient population distributions and realistic clinical diagnosis challenges, than any previous chest X-Ray datasets. The dataset comes with annotated metadata info, consisting of several risk associated factors.

A significant limitation of this dataset is the labeling policy, which may raise some concerns. More specifically, the medical reports were analyzed by an automatic text-mining model, which assigned the corresponding labels according to its text-mining procedure. However, as the authors claim, "there would be some erroneous labels, but the Natural Language Processing (NLP) labeling accuracy is estimated to be >90%" [10].

For the particular experiment, the following disease cases were selected: a) Pulmonary Edema, b) Pleural effusion, c) Chronic obstructive pulmonary disease, d) Pulmonary fibrosis. The selection was based on the significance and frequency of those diseases. Analytically, 293 images representing the Pulmonary Edema, 311 images representing the Pleural Effusion, 315 images representing the Chronic Obstructive Pulmonary Disease (COPD), and 280 images representing the Pulmonary Fibrosis were randomly chosen from the collection.

2.1.4 Image Pre-Processing and Data Augmentation

The X-Ray images were rescaled to a size of 200x200. For the images of different pixel ratios, and to avoid distortion, black background of 200x200 pixels was added to achieve a complete transformation. Low contrast images or images containing parts of the whole thoracic X-Ray scan were excluded.

During the training process, slight augmentations were applied to the images. Data augmentation is mandatory to generate the necessary diversity aiding to the generalization capabilities of the CNNs [11]. Specifically, the images are randomly rotated by a maximum of 10º and randomly shifted horizontally or vertically by a maximum of 20 pixels towards any direction. In this way, the CNN learns to be robust to position and orientation variance.

2.1.5 Data Limitations

The present collection of data faces some limitations, which have to be mentioned. Firstly, a relatively small sample of COVID-19-infected cases is incorporated. Besides, this sample may derive from patient cases with severe symptoms, the analysis of which was mandatory. Cases with slight symptoms are missing from the current public collections, which is due to the policy of protecting people (and society) who have mild symptoms of the disease, and are immediately quarantined without further examination.

Secondly, the pneumonia incidence samples are older recorded samples and do not represent pneumonia images from patients with suspected Coronavirus symptoms, while the clinical conditions are missing.

Thirdly, further data related to demographic characteristics and other potential predisposing or risk factors are not available, and this impedes a holistic approach and examination beyond the medical image.

### 2.2 Learning Strategies for Feature Extraction

There are currently major techniques that successfully employ CNNs to medical image classification, by extracting features, as follows: a) training the CNN from scratch, b) employing a pre-trained CNN, which is called Transfer Learning [12], and c) a hybrid method, which also a Transfer Learning method, and adopts the former strategies by tuning the trainability of specific layers of the CNN; this method is called Fine-Tuning [13].

The first strategy may be adopted either by developing a novel CNN architecture, or by employing the architecture of a successful CNN. In this study, we employ a state-of-the-art CNN architecture, to follow each of the strategies.

The second and the third strategy are means of Transfer Learning. Transfer learning is a machine learning method wherein a model developed for a specific task is reused for another task. There are two categories to perform the Transfer Learning, i.e. the of-the-self feature extraction and Fine-Tuning.

The off-the-self strategy is an approach utilizing the weights of the Convolutional layers, which are defined from the source task (the initial training of another domain) without re-training the network [14]. Extracting such features is usually fast and this approach requires only the addition of a classifier to perform the classification of those features with respect to their significance in the particular task.

The Fine-Tuning strategy involves utilizing a network initialized with pre-trained weights and partially re-training it on the target task. In the context of deep learning, fine-tuning a deep network is a common strategy to learn both task-specific deep features, and retain the methodology to extract global features met in every image, such as shapes. Usually, the Fine-Tuning strategy allows more trainable weights at the top of the network (i.e., the final steps), due to the fact that those convolutional layers extract more abstract and high-level information, compared to the first layers wherein local features are learned. In the particular experiment, we gradually allow more layers to be trainable, by defining 6 experimental cases, referred to as Fine-Tuning (e.g., 11), where the number in the parenthesis corresponds to the number of trainable convolutional blocks.

### 2.3 Method for the extraction of possibly significant biomarkers

#### 2.3.1 The state-of-the-art CNN called Mobile Net

For the classification task, the state-of-the-art CNN called Mobile Net [15] was employed. Mobile Net has been recently utilized for the same classification task by Apostolopoulos [13]. In their work, the authors demonstrated the superiority of Mobile Net in reducing the False Negatives for the detection of COVID-19, compared to other famous CNNs. Besides, this CNN introduces a fewer number of parameters compared to other CNNs, which makes it appropriate for swift training.

The MobileNet [15] model is based on depthwise separable convolutions [16], which is a form of convolutions transforming a conventional convolution into a depthwise convolution [16] and a 1x1 convolution, which is commonly known pointwise convolution [16]. This procedure reduces the number of parameters drastically.

To the top of the Mobile Net v2, a Global Average Pooling [17] layer was added, which drastically reduces the issue of overfitting [18]. The extracted image features are inserted into a Neural Network of 2500 nodes to distinguish between the irrelevant and the significant ones. To further aid to the overfitting reduction, the weights of each feature are normalized utilizing a Batch Normalization layer [19], while we independently zero out the 50% of the outputs of neurons at random, via a Dropout layer [20]. An overview of the method is illustrated in Figure 1.

The intention of the particular study is not only to achieve a high classification accuracy but to achieve this by training the CNN from scratch. This strategy is preferable to transfer learning to evaluate the significance of the features extracted from the precise images, while not depending on features already learned by the pre-trained model, the initial training of which was performed utilizing non-medical images.

Based on the results, the extracted features may be evaluated to conclude that they may constitute real biomarkers related to various diseases.

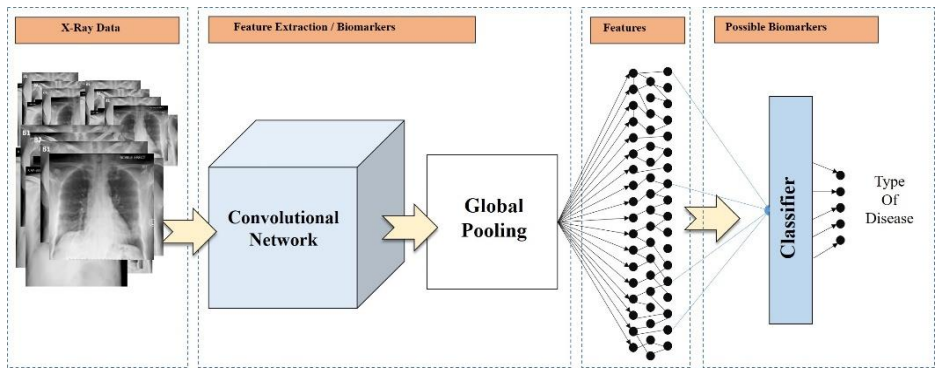

**Fig. 1.** Overview of the feature extraction process

### 2.4 Experiment Setup

We performed a set of three different experiments employing the same CNN (Mobile Net v2) but altering the learning strategy. The following strategies are evaluated: a) Transfer Learning with of-the-self features, b) Transfer Learning with Fine-Tuning, and c) Training from scratch, which, in this experiment, is a latent form of the Transfer Learning, since we only borrow the architecture of the Mobile Net and not the learned parameters. The experiments were performed utilizing a single GPU

8setup (NVIDIA GeForce RTX 2060 Super) using the Keras library [21] and TensorFlow [22] as backend.

The training and evaluation procedure was performed with 10-fold-cross-validation. During this procedure, the dataset is randomly split to ten folds, nine of which are utilized for training the model, and the remaining fold is hidden and used to test the performance and the confidence of the predictions after the training. This process is repeated in a way that every fold is utilized as the test set. This increases the computational cost but enhances the significance of the result. The final accuracy is obtained by calculating the mean accuracy derived from each testing fold.

### 2.5 Metrics

The metrics, based upon which the evaluation of the performance is made, are the overall 7-class accuracy, and the accuracy corresponding to the 2-class classification (COVID-19 vs. non-COVID-19).

Besides, to focus on the performance of COVID-19 detection, the following values are recorded: a) Correctly predicted COVID-19 cases (True Positives), b) Correctly predicted non-COVID-19 cases (True Negatives), c) Incorrectly predicted COVID-19 cases (False Positives), and d) Incorrectly predicted non- COVID-19 cases (False Negatives). Based on those values, the Sensitivity and Specificity of the test are calculated by the following equations:

$$Sensitivity = \frac{True\ Positives}{True\ Positives + False\ Negatives} \quad (1)$$

$$Specificity = \frac{True\ Negatives}{True\ Negatives + False\ Positives} \quad (2)$$

For the particular experiments and given that there is a class imbalance issue, the most reliable metric is that of the 7-class accuracy, while given that this accuracy is high, the second most vital metric is that of Specificity. This is due to the importance of correctly identifying the actual non-COVID-19 cases (True Negatives).

## 3 Results

In this section, the results for the different experiment setups are presented. Based on those results, the optimal strategy is selected, and assumptions are made regarding its effectiveness.

3.1. Results of the Of-the-Self-features strategy

In Table 1, the accuracy, sensitivity, and Specificity of the first strategy are given. The reader should recall that the 2-class accuracy refers to the case where the labels are "COVID-19" and "Non-COVID-19". Besides, the sensitivity and the Specificity refer to the 2-class measurement.

**Table 1.** Accuracy, Sensitivity, and Specificity for the Of-the-Self-features strategy

| Strategy | Accuracy 2-class (%) | Accuracy 7-class (%) | Sensitivity (%) | Specificity (%) |
|---|---|---|---|---|
| Of-the-self-features | 88.81 | 51.98 | 04.62 | - |

Due to the class imbalance, the metric of Specificity was approaching 100% and was not mentioned in Table 1, as it is not a meaningful measurement when reaching those values. The same issue is valid for the 2-class accuracy but was mentioned in Table 1 for comparisons. The confusion matrix for each class is presented in Table 2.

**Table 2.** Confusion Matrix for the 7-class classification employing transfer learning with of-the-self features

|  | Actual Classes | | | | | | |
|---|---|---|---|---|---|---|---|
|  | Covid19 | Edema | Effusion | Emphys. | Fibrosis | Pneumonia | Normal |
| Covid19 | 21 | 0 | 1 | 1 | 0 | 1 | 0 |
| Edema | 270 | 254 | 210 | 199 | 155 | 171 | 136 |
| Effusion | 4 | 5 | 24 | 4 | 6 | 0 | 1 |
| Emphys. | 15 | 16 | 34 | 49 | 31 | 4 | 7 |
| Fibrosis | 46 | 17 | 35 | 50 | 78 | 3 | 18 |
| Pneumonia | 91 | 1 | 3 | 4 | 2 | 712 | 287 |
| Normal | 8 | 0 | 4 | 8 | 8 | 19 | 892 |

(Predicted Classes on rows)

The confusion matrix corresponding to the COVID-19 class vs. all the classes is presented in Table 3.

**Table 3.** Confusion Matrix for the 2-class classification employing transfer learning with of-the-self features

|  | Actual Class | |
|---|---|---|
|  | COVID-19 | Non-COVID-19 |
| **Predicted COVID-19** | 21 | 3 |
| **Predicted Non-COVID-19** | 434 | 3447 |

Based on the results, it is confirmed that the particular strategy is not effective in extracting useful features to distinguish possible underlying information from the X-Rays related to the COVID-19 disease. Besides, a bias towards the non-COVID-19 cases is observed in Table 3, which makes the strategy not appropriate for the particular task.

3.2. Results of the Fine-Tuning strategy

In Table 4, the accuracy, the Sensitivity, and the Specificity of the second strategy are given. The reader should recall that several adjustments for fine-tuning are tested in the particular section, which are discussed in Section 2.2. The number defining each experimental case, refers to the number of blocks made trainable during the experiment, e.g., "Fine-Tuning 3" corresponds to 3 trainable blocks starting from the top of the CNN. The reader should also recall that the values for the accuracy are the mean values of the accuracies obtained at each fold from the 10-fold-cross-validation procedure.

**Table 4.** Accuracy, Sensitivity, and Specificity for the different cases of the Fine-Tuning strategy

| Strategy | Accuracy 2-class (%) | Accuracy 7-class (%) | Sensitivity (%) | Specificity (%) |
|---|---|---|---|---|
| Fine-Tuning 1 | 86.44 | 50.24 | 11.22 | - |
| Fine-Tuning 3 | 88.02 | 56.91 | 57.26 | - |
| Fine-Tuning 5 | 87.66 | 52.67 | 58.08 | - |
| Fine-Tuning 7 | 90.37 | 43.43 | 63.52 | 93.91 |
| Fine-Tuning 9 | 91.28 | 66.31 | 71.84 | 94.55 |
| Fine-Tuning 11 | **92.33** | **75.67** | **82.96** | **93.96** |

As it is observed in Table 4, the strategy of fine-tuning obtains different results. This is explained by the fact that we gradually allow more layers to be trainable, thus approaching close to the strategy of training from scratch, which obtains the best results, as it is presented in Section 3.3. Hence, the confusion matrixes are not provided due to insignificance and limitations of space.

3.3. Results of the Training-from-scratch strategy

In Table 5, the accuracy, Sensitivity, and Specificity of the specific strategy are presented.

**Table 5.** Accuracy, Sensitivity, and Specificity when the training from scratch strategy was followed

| Strategy | Accuracy 2-class (%) | Accuracy 7-class (%) | Sensitivity (%) | Specificity (%) |
|---|---|---|---|---|
| Training from Scratch | 99.18% | 87.66% | 97.36% | 99.42% |

In Table 5, it is observed that training from scratch outperforms the other strategies in terms of every metric, obtaining a remarkable 2-class accuracy of 99.18% and a high 7-class accuracy of 86.66%. The reader should recall that 2-class accuracy refers to the effectiveness of distinguishing the COVID-19 cases from every other case, including both abnormal and normal cases. In Table 6, the confusion matrix for the 7-class task is presented.

**Table 6.** Confusion Matrix for the 7-class classification employing the strategy of training Mobile Net v2 from scratch

|  |  | Actual Classes | | | | | | |
|---|---|---|---|---|---|---|---|---|
|  |  | Covid19 | Edema | Effusion | Emphys. | Fibrosis | Pneumonia | Normal |
| **Predicted Classes** | Covid19 | 443 | 1 | 4 | 4 | 7 | 3 | 1 |
|  | Edema | 1 | 232 | 36 | 34 | 11 | 0 | 0 |
|  | Effusion | 2 | 31 | 161 | 58 | 37 | 0 | 0 |
|  | Emphys. | 3 | 12 | 54 | 156 | 40 | 0 | 0 |
|  | Fibrosis | 3 | 17 | 56 | 63 | 184 | 0 | 0 |
|  | Pneumonia | 1 | 0 | 0 | 0 | 1 | 907 | 0 |
|  | Normal | 2 | 0 | 0 | 0 | 0 | 0 | 1340 |

Several outcomes are to be highlighted in Table 6. Firstly, out of 455 COVID-19 cases, 443 cases were correctly identified, while only 2 cases were mistakenly classified as normal. Secondly, out of the 1341 normal cases, only 1 case was mistakenly identified. For the rest of the pulmonary abnormalities, there is a diversity, which may derive from the fact that the different pathogens embody seals that are difficult to distinguish from the X-Rays and confuse the CNN.

The confusion matrix corresponding to the COVID-19 class vs. all the classes is presented in Table 7.

**Table 7.** Confusion Matrix for the 2-class classification employing the strategy of training Mobile Net v2 from scratch

|  | Actual Class | |
|---|---|---|
|  | COVID-19 | Non-COVID-19 |
| **Predicted COVID-19** | 443 | 20 |
| **Predicted Non-COVID-19** | 12 | 3430 |

The classification obtains an excellent trade-off between the corresponding True Positives, False Positives, True Negatives, and False Negatives.

## 4 Discussion

The particular research focuses on discovering possible biomarkers from X-Ray images. These biomarkers may be significantly related to the COVID-19 disease.

While Deep Learning extracts a massive amount of high-dimensional features from images, it is possible that some of those features behave as actual biomarkers. The reader may be confused by the difference between a feature and a biomarker. Therefore we briefly describe the difference between them. A feature is a specific characteristic of an image, either well-defined in the literature or yet to be defined as to its importance. With Deep Learning, it is possible to extract millions of related features. The extracted features' importance to the specific task is questionable. The majority of those features may be irrelevant to the desired outcome, or the desired subject of study and are rejected by the automatic classification performed after the convolutional layers of a CNN. The biomarkers are quantitative markers of confirmed significance and are not limited to the image features [23]. Generally, the ability of Deep Learning for biomarker extraction is questionable due to the issue of the interpretability.

This study suggests that it may be possible to discover new reliable biomarkers from X-Ray images due to the fact that a high classification accuracy was achieved. Since the CNNs and the Neural Networks lay on the evaluation of millions of parameters to classify the significant features, some of those features may actually be biomarkers leading to a reliable result. This horizon in to be investigated in future research, possibly exploring other approaches, such as Radiomics [24].

One factor that underpins the conclusion mentioned above is the comparison between the various image feature mining strategies. In particular, it is demonstrated that those strategies do not mine the same features. This can be easily interpreted, since with strategies of Transfer Learning with of-the-self-features and Transfer Learning with fine-tuning, the ability of the CNN to extract significant features depends on factors related to the initial training. The initial training was mandatory to be performed on images of a completely different nature due to the absence of large-scale data. However, despite the fact that the latter strategies have excellent performances in other medical image classification tasks [12], [25], in the particular experiment, they were underperforming. This may suggest that with the training from scratch, essential features related to the pulmonary abnormalities have been mined, which may constitute relevant Biomarkers.

In future studies, some issues of the present study can be circumvented. A more in-depth analysis, in particular, requires much more patient data, particularly those suffering from COVID-19.

A more promising approach for future studies would concentrate on identifying patients infected by COVID-19, but showing mild symptoms, although those symptoms may not be visualized correctly on x-rays, or may not be visualized at all.

It is of vital importance to establish models capable of distinguishing between a more significant number of pulmonary diseases, possibly including that of SARS. Also, despite the fact that the appropriate treatment is not determined solely from an X-Ray image [26], an initial screening of the cases would be useful, not in the type of treatment, but in the timely application of quarantine measures in the positive samples, until a complete examination and specific treatment or follow-up procedure are followed.

## 5      Conclusion

The contribution of this work is two-fold. Firstly, low-cost, rapid, and automatic detection of the COVID-19 disease was achieved, utilizing a significantly large sample of several pulmonary infections. Secondly, the study suggests that future research should be conducted to investigate the possible behavior of the extracted feature as Biomarkers since there is sufficient evidence, based on the particular results. Besides, the advantage of automatic detection of COVID-19 from either medical image lies in the reduction of exposure of nursing and medical staff to the outbreak.